\documentclass[11pt,a4paper]{article}
\pdfoutput=1
\usepackage{amsmath,amssymb,tikz}
\usepackage{graphicx}
 \allowdisplaybreaks
\long\def\symbolfootnote[#1]#2{\begingroup%
\def\thefootnote{\fnsymbol{footnote}}\footnote[#1]{#2}\endgroup}
\newcommand{\sysu}{{\it School of Physics and Astronomy, Sun Yat-Sen University, 2 Daxue Road, Zhuhai 519082, China}}

\newcommand{\scnu}{{\it Guangdong Provincial Key Laboratory of Nanophotonic Functional Materials and Devices, South China Normal University, Guangzhou, 510631, China}}

\begin{document}
%%%%%%%%%%%%%%%%%%%%%%%%%%%%%%%%%%%%%%%%%%%
\thispagestyle{empty}
%%%%%%%%%%%%%%%%%%%%%%%%%%%%%%%%%%%%%%%%%%%
%%%%%%%%%%%%%%%%%%%%%%%%%%%%%%%%%%%%%%%%%%%
%\begin{flushright}
%\hfill{AEI-2015-xxx}
%\hfill{ NCTS-TH/1702}
%\end{flushright}
%%%%%%%%%%%%%%%%%%%%%%%%%%%%%%%%%%%%%%%%%%%
\begin{center}

  ~\vspace{20pt}

  {\Large\bf General Junction Condition and Casimir Effect for (1+1)-Dimensional Scalar Network CFT}

  \vspace{25pt}

Tian-Ming Zhao ${}^{1}$ \symbolfootnote[1]{Email:~\sf zhaotm@scnu.edu.cn}, Sinan Pang ${}^{1}$, Ling Li ${}^{1}$, Yu Guo${}^{2}$,  Rong-Xin Miao ${}^{2}$\symbolfootnote[2]{Email:~\sf miaorx@mail.sysu.edu.cn}

  \vspace{10pt}${{}^1}$\scnu

  \vspace{10pt}${{}^2}$\sysu

  \vspace{2cm}

 \begin{abstract}
Recently, BCFT and ICFT have been generalized to the CFT on networks (NCFT). A key aspect of NCFT is how we connect the CFTs across different edges at the network nodes. Previous research has primarily concentrated on a specific junction condition (JC) that requires the field to be continuous at the nodes. In this paper, we investigate the most general junction conditions for $(1+1)$-dimensional free scalars that are consistent with the variational principle and energy conservation. These general junction conditions are characterized by an \(O(p)\) group, where \(p\) represents the number of edges connected at a node. We provide exact realizations of two typical JCs in real physical systems. Additionally, we derive both the lower and upper bounds on the network Casimir energy for $(1+1)$-dimensional free scalar fields and extend the lower bound to encompass general NCFTs. Finally, we analyze the Casimir effect in networks composed of regular polyhedra and examine the binding energy required to construct such networks from individual components. 
\end{abstract}

\end{center}

%%%%%%%%%%%%%%%%%%%%%%%%%%%%%%%
\newpage
\setcounter{footnote}{0}
\setcounter{page}{1}
%%%%%%%%%%%%%%%%%%%%%%%%%%%%%%%

\tableofcontents
%%%%%%%%%%%%%%%%%%%%%%%%%%%%%%%

\section{Introduction}

\begin{figure}[t]
  \centering
  \includegraphics[width=10cm]{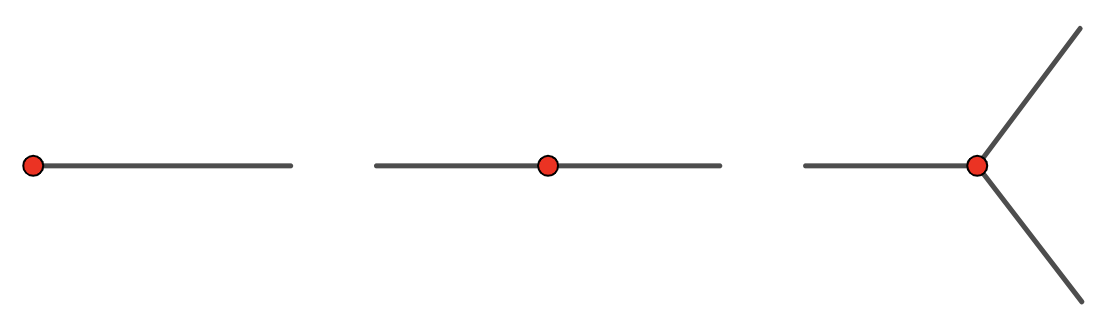}
  \caption{Geometry for BCFT, ICFT and NCFT.}
  \label{NCFT}
\end{figure}

The conformal field theory on networks (NCFT) \cite{Zhao:2025pku, Guo:2025sbm} is a multi-branch generalization of boundary conformal field theory (BCFT) and interface conformal field theory (ICFT). Typically, we have one node with one edge for a BCFT, one node with two edges for an ICFT, and multiple nodes and edges for an NCFT. See Fig. \ref{NCFT} for examples. BCFT and its gravity dual \cite{Takayanagi:2011zk} are powerful tools for studying boundary effects, such as boundary anomalous transports \cite{Chu:2018ksb, Chu:2018ntx, Zhao:2023rno}. They also play a significant role in the recent breakthroughs concerning the black hole information paradox \cite{Penington:2019npb, Almheiri:2019psf, Almheiri:2020cfm}. Therefore, there is a strong motivation to investigate its generalization, the NCFT. NCFT can describe the electron/phonon physics in nanoscale circuits and exhibit new quantum phenomena compared with BCFT. For example, consider the Casimir effect \cite{Casimir:1948dh, Plunien:1986ca, Bordag:2001qi, Milton:2004ya, Bordag:2009zz}. In BCFT, the Casimir force is always attractive when the same boundary conditions are imposed on two planes \cite{Bachas:2006ti}. This attractive force is harmful in nanotechnology, potentially damaging delicate structures and restricting the movement of nano-devices \cite{Chan1, Chan2}. However, with NCFT, the Casimir force can be adjusted from attractive to repulsive by varying the lengths of the network edges, providing a practical means to control the Casimir effect \cite{Zhao:2025pku}. 

\begin{figure}[t]
  \centering
  \includegraphics[width=7.66cm]{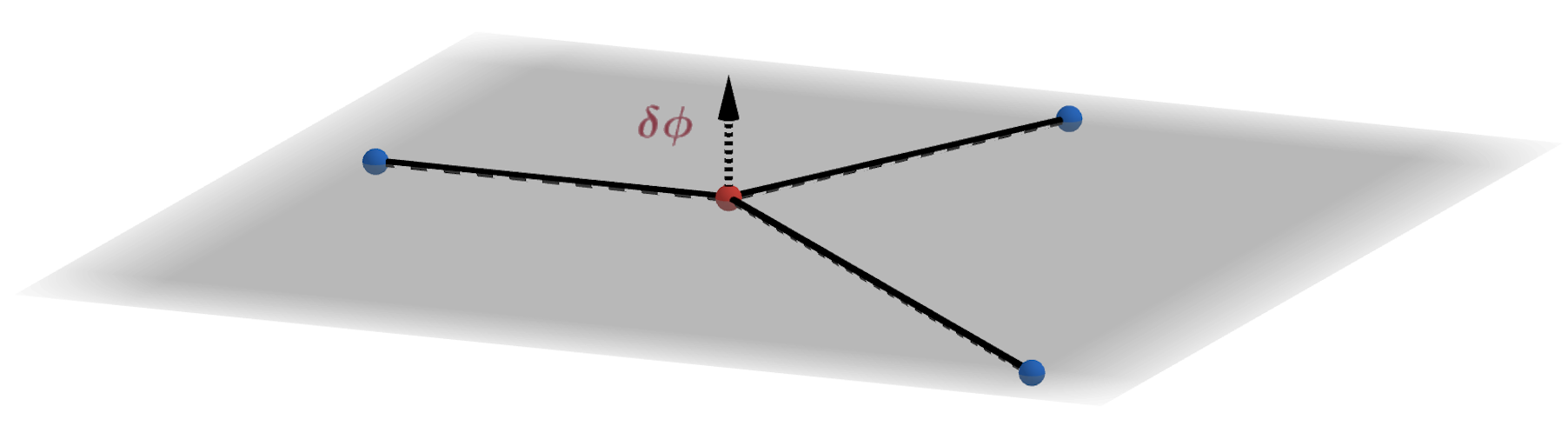} \includegraphics[width=7.66cm]{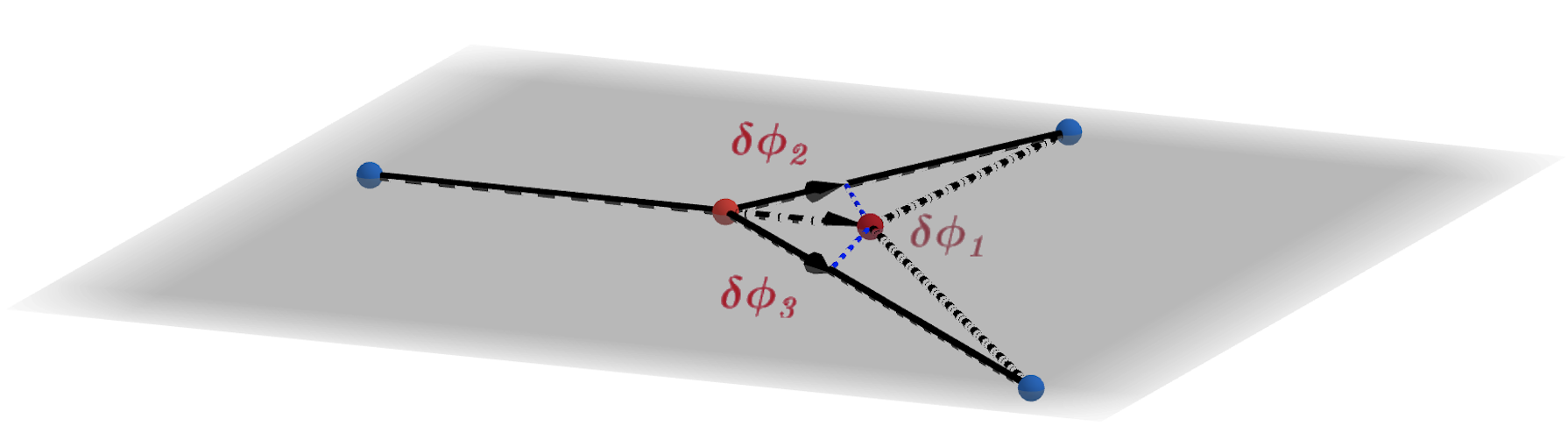}
  \caption{(Left) Transverse vibration of three linked strings in a plane; (Right) Longitudinal vibration of three linked rigid rods in a plane. We have $\delta \phi_1|_N=\delta \phi_2|_N=\delta \phi_3|_N$ and $\delta (\phi_1+\phi_2+\phi_3)|_N=0$ on the nodes $N$ (red points) for left and right figures. The arrows denote the micro-displacements $\delta \phi_i$. }
  \label{realnetworks}
\end{figure}

The junction condition (JC) plays a central role in NCFT, as it determines how to connect the CFTs in different edges at the nodes of the network. Take $(1+1)$-dimensional free scalar as an example, the corresponding JC is given by \cite{Zhao:2025pku}
\begin{eqnarray} \label{JC1}
  \text{JC I}: \ \phi_i|_N=\phi_j|_N,\  \ \sum_{i=1}^p \partial_{n} \phi_i|_N=0,
\end{eqnarray}
where $\phi_i$ denotes the scalar on edge $E_i$, $p$ is the numbers of edges linked by the node $N$, and $n$ labels the inward-pointing normal vector on the node $N$. The JC I (\ref{JC1}) is derived from the variational principle together with the requirement that the scalar fields are continuous at the nodes, i.e., $\phi_i|_N=\phi_j|_N$. It is consistent with the conservation of energy at nodes \cite{Zhao:2025pku}
\begin{eqnarray} \label{conservation law}
  \sum_{i=1}^p \overset{(i)}{T}_{nt}|_N= \sum_{i=1}^p  \partial_n \phi_i \partial_t \phi_i |_N=0,
\end{eqnarray}
that the total energy flux flowing into the node is zero. We observe that JC I (\ref{JC1}) is a sufficient but not necessary condition for the conservation of energy (\ref{conservation law}). In fact, the conservation law (\ref{conservation law}) also holds under the following alternative JC
\begin{eqnarray} \label{JC2}
  \text{JC II}: \ \partial_{n}\phi_i|_N=\partial_{n}\phi_j|_N,\  \ \sum_{i=1}^p  \phi_i|_N=0.
\end{eqnarray}
It can be derived from the variational principle by requiring the normal derivative of the scalar field to be continuous at the nodes, i.e., $\partial_{n}\phi_i|_N=\partial_{n}\phi_j|_N$. See sect. 2.1 for more details. It seems odd to require a continuous normal derivative of the field at the nodes. To clarify this concern, we present a clear example of JC II using a real physical network composed of thin rigid rods. See Fig. \ref{realnetworks} (right). In this setup, the scalar field represents the longitudinal micro-displacement of the rods. As illustrated, when one rod expands by $\phi_i>0$, it compresses the other rods ($\phi_j<0$, where $j \ne i$), which leads to the second equation of JC II (\ref{JC2}). The conservation law (\ref{conservation law}) or the variational principle then gives us the first equation of JC II (\ref{JC2}). For more details, see sect. 2.1. Conversely, JC I can be demonstrated with a network of strings, where the scalar represents the transverse micro-displacement. In this case, we naturally have $\phi_i|_N = \phi_j|_N$ as shown in Fig. \ref{realnetworks} (left).

In addition to JC I and JC II, there exist more general JCs that comply with the conservation law (\ref{conservation law}). The second term in (\ref{conservation law}) indicates an \( O(p) \) rotation symmetry for \( \phi_i \), which leads to these broader JCs. For further details, please refer to (\ref{general JC solution2}) in sect. 2.2. We examined the Casimir effect in the simplest network, consisting of \(p\) edges connected at a single node, focusing on these general JCs. We discovered both lower and upper bounds on the network's Casimir energy for free scalars in $(1+1)$ dimensions. Notably, these bounds are saturated by the JCs that reduce to the perfectly reflecting boundary conditions in BCFTs. Inspired by the recent work \cite{Miao:2024gcq, Miao:2025utb}, we propose that the lower bound of the Casimir effect can be extended to general NCFTs in various networks, beyond just free scalars in the simplest configuration:
\begin{eqnarray} \label{network Casimir bound}
 W\ge -\frac{\pi c}{24}\sum_{i}^p \frac{1}{L_i},
\end{eqnarray}
where $c$ is the central charge, $L_i$ is the length of edge $E_i$ and the sum is taken over all edges. More discussions can be found in sect. 3.

In a previous study \cite{Zhao:2025pku}, two authors of this paper examined the planar network with JC I. This paper further explores the Casimir effect in spatial networks with both JC I and JC II, particularly in symmetric networks composed of regular polyhedra. The case of general JCs is complicated for regular polyhedra, and we will address it in future work. We introduce the concept of network binding energy, defined as the Casimir energy of the network minus the Casimir energy of its isolated constituent units (strips):
\begin{eqnarray} \label{network binding energy}
 W_{\text{bind}}=W_{\text{NCFT}}-W_{\text{BCFT}}.
 \end{eqnarray}
Different boundary condition choices for the strip in BCFTs yield different Casimir energies. If we impose the same boundary conditions on the two boundaries of the strip, we get a minimal Casimir energy given by $W_i=-\pi c/(24 L_i)$ \cite{Miao:2024gcq}. Consequently, due to the bound (\ref{network Casimir bound}), the network binding energy is non-negative under this choice. It implies that constructing a network from isolated strips with identical boundary conditions requires energy. The Casimir effect of regular polyhedra with unit edge length is summarized in Table \ref{table1}, where $\rho$ represents the Casimir energy density, and $\rho_{\text{bind}}= W_{\text{bind}}/(\sum_i L_i)$ is the binding energy density. The negative $\rho$ indicates that the Casimir force is attractive for both JC I and JC II. Additionally, the Casimir force for JC II is usually not greater than for JC I. Notably, for the Hexahedron network, both JC I and JC II produce the same Casimir force. The positive value of \( \rho_{\text{bind}} \) in brackets further supports our earlier assertion that the network binding energy is non-negative under our choice. 
\begin{table}[h]
  \centering
  \caption{Casimir effect of regular polyhedra with unit edge length}
  \begin{tabular}{|c|c|c|c|c|c|}
    \hline
    $\rho \  (\rho_{\text{bind}})$    & Tetrahedron & Hexahedron & Octahedron & Dodecahedron & Icosahedron \\
    \hline
    JC I  & -0.06 (0.07)       & -0.04 (0.09)      & -0.06 (0.07)     & -0.03 (0.10)        & -0.07 (0.06)     \\
    \hline
    JC II & -0.01  (0.12)     & -0.04  (0.09)    & -0.04 (0.09)    & -0.02  (0.11)      & -0.06 (0.07)      \\
    \hline
  \end{tabular} \label{table1}
\end{table}

The paper is structured as follows: In Section 2, we examine the general junction conditions (JCs) for free scalars in $(1+1)$ dimensions. We begin by analyzing two typical JCs and provide exact implementations of them in a real physical system. Next, we explore the most general JCs that are consistent with energy conservation and express them in terms of elements from the \(O(p)\) group. Section 3 investigates the bounds of network Casimir energy. We start with the simplest network as an example and then argue that our results hold for general networks. In Section 4, we analyze the Casimir effect on networks composed of regular polyhedra, applying both JC I and JC II. Finally, we conclude with a discussion in Section 5. 
% v2 The appendix provides the spectra of various networks. 
Appendix A analyzes the JCs for the scalar field and the Maxwell field in general dimensions. Appendix B presents the spectra of various networks.
Note that we take the natural units with $c=\hbar=1$ in this paper.

\section{General Junction Conditions}

% v2
This section derives the general junction conditions (JCs) consistent with the action principle and energy conservation at nodes, as stated in equation (\ref{conservation law}), for (1+1)-dimensional free scalar fields. To begin, we will first examine two typical JCs derived from the action principle and their physical realizations in the real world. Subsequently, we will derive the most general junction conditions based on energy conservation at the nodes. In this paper, we focus on free scalars in (1+1) dimensions, but generalizations to other field theories and higher dimensions are straightforward. For more details, please refer to Appendix A.

\subsection{Typical JCs}

Let us start with the action of a 2d free scalar
\begin{eqnarray}\label{action}
  I= \frac{1}{2} \sum_{i=1}^p \int_{E_i} dt dx_i \Big( \partial_t\phi_i \partial_t\phi_i-\partial_{x_i }\phi_i \partial_{x_i }\phi_i \Big),
\end{eqnarray}
where $\phi_i$ represents the scalar field on the edge $E_i$, and the edges $E_i$ intersect at the node $N$ with $x_i = 0$.
Taking variations of the action yields boundary terms at the node
\begin{eqnarray} \label{action1}
  \delta I|_N= \int_{N} dt  \Big( \sum_{i=1}^p \partial_{n} \phi_i \delta \phi_i \Big)=0,
\end{eqnarray}
where $n$ denotes the normal vector pointing from node $N$ to edge $E_i$. For a well-defined variational principle, we need $\delta I|_N$ to vanish. We have two natural options to achieve this condition. The first option is to require that the scalar field is continuous at the node, i.e., $ \phi_i|_N=\phi(t)$. Additionally, we require that the scalar field is dynamical at the node, meaning $\delta \phi|_N \ne 0$. Otherwise, we get a BCFT instead of a non-trivial NCFT. Under this assumption, the action variation (\ref{action1}) becomes
\begin{eqnarray} \label{action2}
  \delta I|_N= \int_{N} dt  \Big(  \sum_{i=1}^p \partial_{n}\phi_i   \Big)\delta \phi=0,
\end{eqnarray}
which yields JC I (\ref{JC1}). It is easy to see that JC I (\ref{JC1}) agrees with the energy conservation at nodes (\ref{conservation law})
\begin{eqnarray} \label{conservation law1}
  \sum_{i=1}^p \overset{(i)}{T}_{nt}|_N= \sum_{i=1}^p  \partial_n \phi_i \partial_t \phi_i |_N= \Big( \sum_{i=1}^p \partial_n \phi_i |_N\Big) \partial_t \phi =0.
\end{eqnarray}
The second option is to require that the normal derivative of the scalar field is continuous at the node $\partial_{n}\phi_i|_N=\partial_{n}\phi$. We similarly require  $\partial_{n}\phi|_N \ne 0$ to establish a non-trivial NCFT rather than a BCFT. From this choice, we obtain
\begin{eqnarray} \label{action3}
  \delta I|_N= \int_{N} dt\  \partial_{n}\phi  \ \delta \Big( \sum_{i=1}^p \phi_i \Big) =0,
\end{eqnarray}
which results in JC II (\ref{JC2}). JC II (\ref{JC2}) is also consistent with the conservation of energy
\begin{eqnarray} \label{conservation law2}
  \sum_{i=1}^p \overset{(i)}{T}_{nt}|_N= \sum_{i=1}^p \partial_n \phi_i  \partial_t \phi_i |_N=  \partial_n \phi_i \partial_t \Big( \sum_{i=1}^p \phi_i |_N\Big)=0.
\end{eqnarray}

To better understand the two JCs, we will provide specific physical realizations of them. As shown in Fig. \ref{realnetworks} (left), consider three strings connected by a node in a plane. Let $\phi$ denote the micro-displacement of the strings perpendicular to the plane. $\phi$ obeys the wave equation
\begin{eqnarray}\label{wave equation}
  \partial_t^2 \phi-v^2\partial_x^2 \phi=0,
\end{eqnarray}
where $v=\sqrt{T/\lambda}$ with $T$ representing the string tension and $\lambda$ representing the linear mass density. Since the strings are linked at the node, the displacements of the strings must be the same at that node, $ \phi_i|_N= \phi_j|_N$. This condition, in conjunction with the action principle (\ref{action2}), leads to JC I (\ref{JC1}).

On the other hand, consider three symmetric rigid rods connected by a node in a plane in Fig. \ref{realnetworks} (right). The longitudinal micro-displacement of the rods also follows the wave equation outlined above. The only difference is that the velocity in this case is given by $v=\sqrt{E/\rho}$, where $E$ is the Young modulus and $\rho$ is the volume mass density. When one rigid rod expands longitudinally $\phi_1$, the other two rigid rods are compressed longitudinally by $\phi_2 =\phi_3 =-\phi_1/2$. There are also induced transverse displacements, but they are irrelevant to our purpose. Hence, we find that $\phi_1+\phi_2 +\phi_3 =0$ for the longitudinal vibration of the first rod.  The same relationship holds for the longitudinal vibrations of the other two rods. Combining the three independent vibrations, we have $\phi_1+\phi_2 +\phi_3 =0$  with $\phi_2$ and $\phi_3$ being free variables. The action variation (\ref{action1}) results in
\begin{eqnarray} \label{action4}
  \delta I|_N= \int_{N} dt \Big[ ( \partial_{n} \phi_2 -\partial_{n} \phi_1 ) \delta \phi_2 + ( \partial_{n} \phi_3 -\partial_{n} \phi_1 ) \delta \phi_3 \Big]=0,
\end{eqnarray}
which leads to JC II, i.e., $\partial_{n} \phi_1=\partial_{n} \phi_2=\partial_{n} \phi_3$.

In summary, we have obtained JC I and JC II from the action principle and  provided strong evidence of them in this subsection.

\subsection{General JCs}

Let us consider more general junction conditions (JCs) that ensure energy conservation at nodes. Note that the conservation law (\ref{conservation law}) is equivalent to the action variation (\ref{action1}) when we set \(\delta \phi_i = \partial_t \phi_i\). We take the method outlined in \cite{Bachas:2001vj} for interface conformal field theories (ICFT) with one node and two edges. Below, we generalize this method to NCFTs with $p$ edges.

We define $v_i = t + x_i$ and $u_i = t - x_i$, and we can rewrite the conservation law (\ref{conservation law}) as follows:
\begin{align}\label{conservation law 3}
    \sum_{i}^{p}\overset{(i)}{T}_{tx}|_{N}=\sum_{i}^{p}\partial_{t}\phi_i \partial_{x}\phi_i|_{N}=\sum_{m}^{p}\Big( (\partial_{v}\phi_i)^2-(\partial_{u}\phi_i)^2 \Big)|_{N}=0.
\end{align}
For simplicity, we have omitted the index $i$ for $x_i$, $u_i$, and $v_i$. By defining a column vector 
\begin{align}
    \Phi=(\phi_1, \phi_2,..., \phi_p)^T,
\end{align}
we can further simplify (\ref{conservation law 3}) as
\begin{align}\label{conservation law 4}
    \partial_{v}\Phi^T \cdot\partial_{v}\Phi=\partial_{u}\Phi^T\cdot\partial_{u}\Phi,
\end{align}
which can be solved as 
\begin{align}\label{general JC solution1}
\partial_{v}\Phi= S\cdot \partial_{u}\Phi, \ \ S^{T}\cdot S=\mathbf{1}_{p\times p}.
\end{align}
Here, $S$ is an element of the $O(p)$ group, and $\mathbf{1}_{p \times p}$ is the identity matrix. Returning to the $(x,t)$ coordinate system, we obtain the general junction condition:
\begin{align}\label{general JC solution2}
(S-\mathbf{1}_{p\times p})\cdot \partial_{t}\Phi=(S+\mathbf{1}_{p\times p})\cdot \partial_{x}\Phi.
\end{align}

Let us consider an example where $p=3$ and express the group element $S$ using Rodrigues' formula:
\begin{align}\label{sect2: S1}
    S=\pm\Big( \cos(\theta)\mathbf{1}_{3\times 3}+(1-\cos(\theta))\hat{n} \hat{n}^T+\sin(\theta)\hat{n}^{\times}\Big),
\end{align}
where $\theta$ is the rotation angle around the rotation axis $\hat{n}$:
\begin{align}\label{sect2: n}
  \hat{n}=(n_x,n_y,n_z)=\Big(\sin(\alpha) \cos(\beta),\sin(\alpha) \sin(\beta),\cos(\alpha) \Big).
\end{align}
Besides, we have
\begin{align}\label{sect2: nn}
\hat{n} \hat{n}^T&=\left(
\begin{array}{ccc}
 n_x n_x & n_x n_y & n_x n_z\\
n_y n_x & n_y n_y & n_y n_z\\\
n_z n_x & n_z n_y & n_z n_z\ \\
\end{array}
\right),
\end{align}
and 
\begin{align}\label{sect2: n times}
    n^{\times}&=\left(
\begin{array}{ccc}
 0 & -n_z & n_y \\
 n_z & 0 & -n_x\\
-n_y & n_x & 0 \\
\end{array}
\right).
\end{align}

Let us discuss some typical cases of the S-matrix and the corresponding JC (\ref{general JC solution2}).  First, when $\theta=0$, we have $S=\pm\mathbf{1}_{p\times p}$, which correspond to NBC and DBC of BCFTs:
\begin{align}\label{sect2: S NBC DBC}
\text{NBC}:   \partial_{x}\Phi|_N=0,\ \ \ \text{DBC}:  \partial_{t}\Phi|_N=0.
\end{align}
Recall that BCFTs satisfy  $\overset{(i)}{T}_{tx}|_{N}=0$, which obey the conservation law $\sum_i\overset{(i)}{T}_{tx}|_{N}=0$ trivially. Second, when $\theta=\pi$, we obtain the following JCs on the node $N$:
\begin{align}
    \partial_{t}\phi_1=\frac{1}{\lambda_{2}}\partial_{t}\phi_2=\frac{1}{\lambda_{3}}\partial_{t}\phi_3,~\partial_{x}\phi_1+\lambda_{2}\partial_{x}\phi_2+\lambda_{3}\partial_{x}\phi_3=0,
\end{align}
and 
\begin{align}
    \partial_{x}\phi_1=\frac{1}{\lambda_{2}}\partial_{x}\phi_2=\frac{1}{\lambda_{3}}\partial_{x}\phi_3,~\partial_{t}\phi_1+\lambda_{2}\partial_{t}\phi_2+\lambda_{3}\partial_{t}\phi_3=0,
\end{align}
which correspond to $+$ and $-$ in the S-matrix (\ref{sect2: S1}), respectively. Here, $\lambda_{2}=\tan(\beta)$ and $\lambda_{3}=\cot (\alpha) \sec (\beta)$.  It is worth noting that the above two JCs reduce to JC I (\ref{JC1}) and JC II (\ref{JC2}) when $\lambda_{2}=\lambda_{3}=1$.

\section{Bounds of network Casimir energy}

In this section, we analyze the network Casimir energy for the general JCs (\ref{general JC solution2}). For simplicity, we concentrate on the simplest network configuration comprising one node and three edges. By using the general JCs, we derive both lower and upper bounds for the Casimir energy. Notably, these bounds correspond to the typical JCs outlined in (\ref{sect2: S NBC DBC}), which result in perfect reflection at the node. In other words, for the JCs (\ref{sect2: S NBC DBC}), all modes are confined to their respective edges and cannot move from one edge to another.

We impose either DBC or NBC at the outer boundaries of edges 
\begin{eqnarray} \label{sect 3:  DBC NBC}
\text{DBC}:\ \phi_i|_{x_i=L_i}, \ \ \text{NBC}:\ \partial_x\phi_i|_{x_i=L_i}=0, \ \ i=1,2,3,
\end{eqnarray}
where $L_i$ represents  the length of edge $E_i$. The scalar fields that comply with these boundary conditions are given by
\begin{eqnarray} \label{sect 3: scalar DBC}
&\text{DBC}:\ \phi_i= a_i \sin\Big(\omega (x_i-L_i)\Big)e^{-i \omega t}, \\
&\text{NBC}:\ \phi_i= a_i \cos\Big(\omega (x_i-L_i)\Big)e^{-i \omega t},  \label{sect 3: scalar NBC}
\end{eqnarray}
where \( a_i \) is a constant, and \( \omega \) represents the frequency to be determined. By substituting these expressions into the general JCs (\ref{general JC solution2}), with \( S \) defined by (\ref{sect2: S1}), we obtain three equations given by \( M_{ij} a_j = 0 \). For non-trivial solutions, where at least one \( a_i \) is non-zero, the determinant of \( M \) must vanish, such that \( |M| = 0 \). This condition enables us to derive the spectrum of \( \omega \). In the symmetric case where \( L_i = L \), we find the spectrum:
\begin{eqnarray} \label{sect 3: spectrum DBC}
&\Delta_{\text{DBC}}^+(\omega)=\Delta_{\text{NBC}}^-(\omega)=\cos (L \omega ) (\cos (\theta )+\cos (2 L \omega )), \\
&\Delta_{\text{DBC}}^-(\omega)=\Delta_{\text{NBC}}^+(\omega)=\sin (L \omega ) (\cos (\theta) -\cos (2 L \omega )),  \label{sect 3: spectrum NBC}
\end{eqnarray}
where $``\pm"$ indicates the choice of \( S \) from (\ref{sect2: S1}). We exclude the zero modes \( \omega = 0 \), as they do not contribute to the Casimir energy. Notably, the spectrum for \( L_i = L \) is independent of the parameters \( \alpha \) and \( \beta \) in the S-matrix. However, for cases with different \( L_i \), the spectrum does indeed depend on \( \alpha \) and \( \beta \). For more details, please refer to the appendix. 

The Casimir energy can be calculated using the formula \cite{Bordag:2001qi}
\begin{eqnarray} \label{key trick}
  W_0=\sum_n \frac{\omega_n}{2}=\frac{1}{2\pi i}\oint \frac{\omega}{2} d \ln \Delta(\omega),
\end{eqnarray}
where the contour integral is taken along the right infinite semicircle and the imaginary axis. In appropriate regularizations, the integral over the infinite semicircle vanishes \cite{Bordag:2001qi}. The integral along the imaginary axis diverges, so we must subtract the contribution from $\omega\to \infty$ (or $L_i\to \infty$) \cite{Bordag:2001qi}. This gives us a finite Casimir energy \cite{Zhao:2025pku}
\begin{eqnarray} \label{Casimir energy}
  W=\int_0^{\infty}\Big( \frac{\omega \sum_{i=1}^p L_i}{2\pi}-\frac{i \omega \Delta '(i \omega)}{2 \pi  \Delta (i \omega)} \Big) d\omega,
\end{eqnarray}
where $p=3$ in our case. Substituting the spectrum (\ref{sect 3: spectrum DBC},\ref{sect 3: spectrum NBC}) into (\ref{Casimir energy}), we obtain the Casimir energy for $L_i=L$: 
\begin{eqnarray} \label{sect 3: energy DBC}
&W_{\text{DBC}}^+=W_{\text{NBC}}^-=\frac{1}{2\pi L}\int_0^{\infty } w \left(3-\frac{\tanh (w) (\cos (\theta )+3 \cosh (2 w)+2)}{\cos (\theta )+\cosh (2 w)}\right) \, dw, \\
&W_{\text{DBC}}^-=W_{\text{NBC}}^+=\frac{1}{2\pi L}\int_0^{\infty } w \left(3-\frac{\coth (w) (\cos (\theta )-3 \cosh (2 w)+2)}{\cos (\theta )-\cosh (2 w)}\right) \, dw. \label{sect 3: energy NBC}
\end{eqnarray}

\begin{figure}[t]
  \centering
  \includegraphics[width=7.5cm]{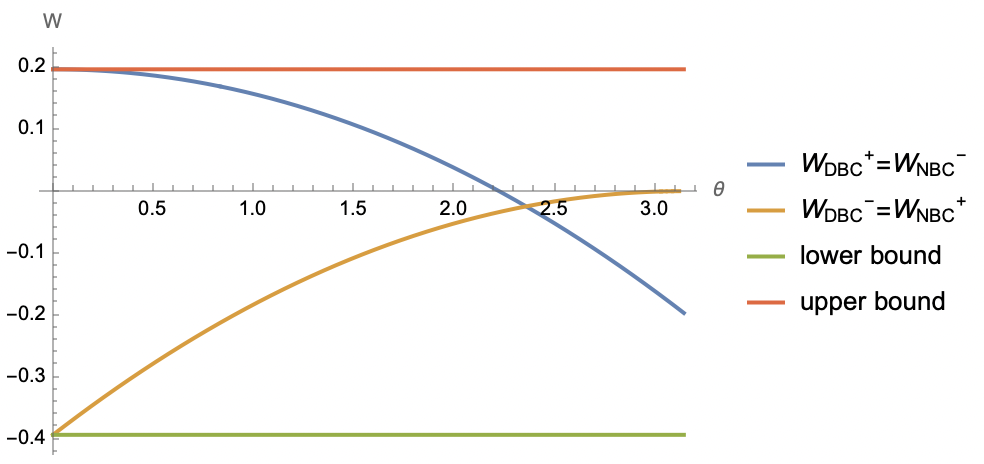}  \includegraphics[width=7.5cm]{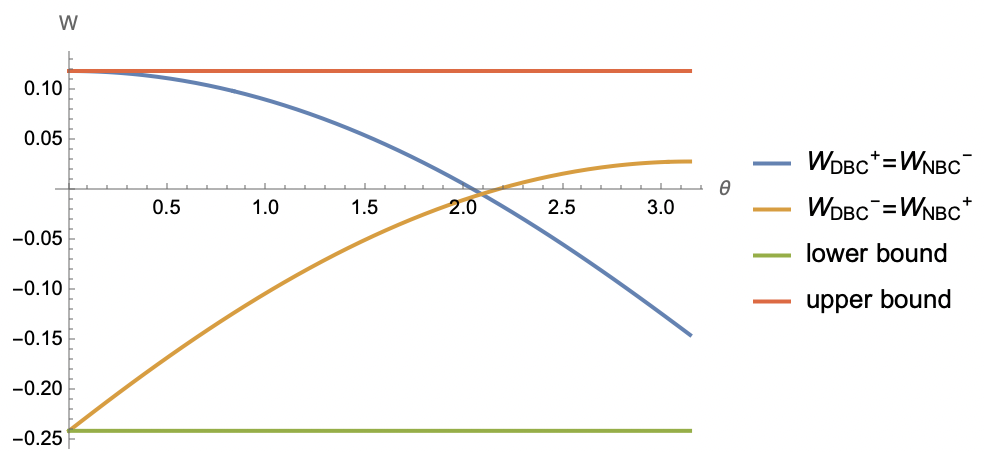}
  \caption{Casimir energy for the simplest network. The left figure is for $L_i=1$ and arbitrary $\alpha, \beta$, while the right figure is for $L_1=1, L_2=2, L_3=3, \alpha=0.3 \pi, \beta=0.6 \pi$.  The two figures both show $W_{\text{DBC}}^+=W_{\text{NBC}}^-$ decreases with the angle $\theta$, while $W_{\text{DBC}}^-=W_{\text{NBC}}^+$ increases with the angle $\theta$ of S-matrix. Besides, these Casimir energy are bounded by the case $\theta=0$, which corresponds to DBC and NBC of BCFTs. }
  \label{Casimir energy 1}
\end{figure}

We illustrate the Casimir energy in Fig. \ref{Casimir energy 1} (left), which shows that \( W_{\text{DBC}}^+ = W_{\text{NBC}}^- \) decreases with the angle \( \theta \), while \( W_{\text{DBC}}^- = W_{\text{NBC}}^+ \) increases with the angle \( \theta \) of the S-matrix. Notably, the Casimir energy is bounded by the case where \( \theta = 0 \), corresponding to perfectly reflecting boundary conditions (DBC and NBC) in BCFTs. We confirm, as illustrated in Fig. \ref{Casimir energy 1} (right), that this holds for general \( L_i \). We have tested a wide range of parameters, not limited to those shown in Fig. \ref{Casimir energy 1}. These figures indicate that the Casimir energy is bounded from below and above by:
\begin{eqnarray} \label{sect 3: Casimir bound}
-\frac{\pi}{24}\sum_{i}^p \frac{1}{L_i}\le W \le \frac{\pi}{48}\sum_{i}^p \frac{1}{L_i},
\end{eqnarray}
where the lower and upper bounds are determined by the Casimir energy of \( p \) isolated strips with identical boundary conditions (NBC-NBC, DBC-DBC) and mixed boundary conditions (NBC-DBC, DBC-NBC), respectively. 

To explain why the Casimir energy of a network is bounded by the JC that reduces to BCFTs, let us recall an important fact: The Casimir effect is a boundary effect, meaning that the higher the reflection coefficient of the boundary, the stronger the Casimir effect \cite{Bachas:2001vj}. The JC with \( \theta = 0 \) reduces to perfectly reflecting boundary conditions in BCFTs, and thus is expected to yield the most significant Casimir effect. It has been proven for general 2d CFTs that the Casimir energy of a strip is bounded from below \cite{Miao:2024gcq}:
\begin{eqnarray} \label{sect 3: strip bound}
 W_{\text{strip}} \ge -\frac{\pi c}{24 L},
\end{eqnarray}
where $c$ is the central charge and $L$ is the strip width. Based on the above arguments, we conclude that the Casimir effect of a network including  \( p \) edges is also bounded from below: 
\begin{eqnarray} \label{sect 3: Casimir bound1}
W\ge -\frac{\pi c}{24}\sum_{i}^p \frac{1}{L_i},
\end{eqnarray}
which aligns with (\ref{sect 3: Casimir bound}) by noting that $c=1$ for free scalars. 

Some comments are in order. First, the arguments supporting the bound presented in (\ref{sect 3: Casimir bound1}) apply to general networks and are not limited to the simplest configuration, which consists of a single node with no loops. Let us quickly test the bound (\ref{sect 3: Casimir bound1}) for a network that includes a loop. Consider two strips with lengths \(L_1\) and \(L_2\) that are joined together to form a loop. The Casimir energy for a loop with periodic boundary conditions is given by $W_{\text{loop}}=-\pi c/6(L_1+L_2)$ \cite{Bordag:2009zz}, which indeed satisfies the bound (\ref{sect 3: Casimir bound1})
\begin{eqnarray} \label{sect 3: test bound}
W_{\text{loop}}=-\frac{\pi c}{6(L_1+L_2)} \ge  -\frac{\pi c}{24}(\frac{1}{L_1}+ \frac{1}{L_2}),
\end{eqnarray}
where the inequality is saturated by $L_1=L_2$. Second, the scalar Casimir effect (\ref{sect 3: Casimir bound}) suggests that there is also an upper bound on the Casimir energy of the network. We will address the upper bound and provide additional tests of the lower bounds in future work.

\section{Casimir effect of polyhedra networks}

\begin{figure}[t]
  \centering
  \includegraphics[width=5cm]{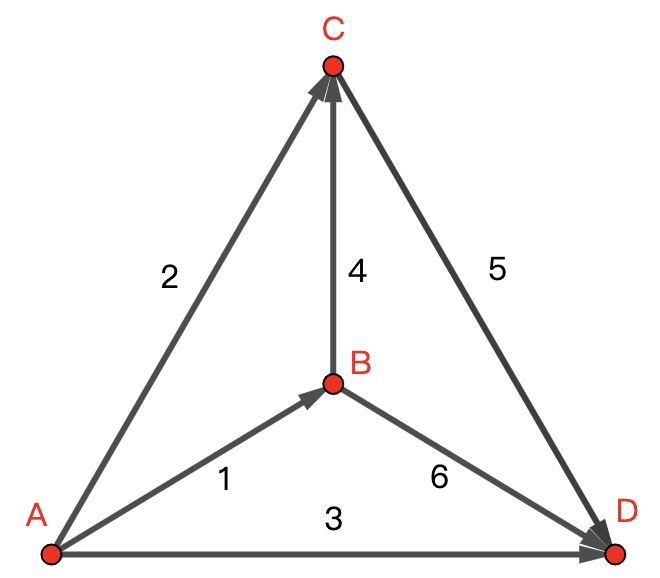}
  \caption{Regular polyhedra with four nodes and six edges. The arrow indicates the direction of the coordinate $x_i $ on edge $E_i$ with $0\le x_i \le L$. Note that the faces do not belong to the network. }
  \label{Tetrahedron}
\end{figure}

This section examines the Casimir effect in networks of regular polyhedra. For simplicity, we focus on JC I (\ref{JC1}) and JC II (\ref{JC2}) and address the general JC (\ref{general JC solution2}) in future work. The numerous nodes and edges in regular polyhedra complicate the discussions of general JCs. To illustrate our method, we use the regular tetrahedron as an example and summarize the results for other regular polyhedra.

The geometry of a regular tetrahedron is illustrated in Fig. \ref{Tetrahedron}. It is important to note that the network is composed of nodes and edges, but does not include the faces of the regular tetrahedron. The arrow indicates the direction of the coordinate $x_i$ along the edge $E_i$, where $0 \leq x_i \leq L$. Denote the scalar field at edge $E_i$ by
\begin{eqnarray} \label{scalar}
  &&\phi_i= \Big( a_i \sin(\omega x_i)+ b_i \sin(\omega x_i) \Big)e^{-i \omega t}, \ i=1,2,..,6
\end{eqnarray}
where $a_i, b_i$ are twelve constants and $\omega$ is the frequency to be determined. It can be verified that (\ref{scalar}) satisfies the wave equation (\ref{wave equation}) with $v=1$. By imposing JCs on the nodes, we derive a set of linear equations. Taking JC II (\ref{JC2}) as examples, we have the following conditions at the nodes
\begin{eqnarray} \label{scalar JC2 AB}
  &&\text{A}:  \ \phi'_1(0)-\phi'_2(0)=0, \  \phi'_2(0)-\phi'_3(0)=0, \   \phi_1(0)+\phi_2(0)+\phi_3(0)=0, \\
  &&\text{B}:  \ -\phi'_1(L)-\phi'_4(0)=0, \  \phi'_4(0)-\phi'_6(0)=0, \   \phi_1(L)+\phi_4(0)+\phi_6(0)=0,\\
  &&\text{C}:  \ -\phi'_2(L)+\phi'_4(L)=0, \  -\phi'_4(L)-\phi_5(0)=0, \   \phi_2(L)+\phi_4(L)+\phi_5(0)=0,\\
  &&\text{D}:  \ -\phi'_3(L)+\phi'_5(L)=0, \  -\phi'_5(L)+\phi'_6(L)=0, \   \phi_3(L)+\phi_5(L)+\phi_6(L)=0,
\end{eqnarray}
where $'$ denote $\partial_{x_i}$. Note that  $\partial_n=\partial_{x_i}$ at $x_i=0$ while $\partial_n=-\partial_{x_i}$ at $x_i=L$. Care should be taken regarding the direction of $x_i$ as indicated by the arrows. Substituting (\ref{scalar}) into the above equations, we get 12 linear equations $ M\cdot A=0$ for 12 unknowns $A=(a_i,b_i)$, where $M$ is a $12\times 12$ matrix. To have non-zero solutions $A$, the determinant of $M$ must vanish $|M|=0$. This approach allows us to determine the spectrum for JC II as
\begin{eqnarray} \label{tetrahedron spectrum II}
  \Delta_{\text{II}}(\omega)=\sin^2\left(\frac{L \omega }{2}\right) \cos ^4\left(\frac{L \omega }{2}\right) (3 \cos (L \omega )-1)^3=0.
\end{eqnarray}
Following the same methodology, we obtain the spectrum for JC I (\ref{JC1})
\begin{eqnarray} \label{tetrahedron spectrum I}
  \Delta_{\text{I}}(\omega)=\sin ^4\left(\frac{L \omega }{2}\right) \cos ^2\left(\frac{L \omega }{2}\right) (3 \cos (L \omega )+1)^3=0.
\end{eqnarray}
Substituting the spectrums (\ref{tetrahedron spectrum I}) and (\ref{tetrahedron spectrum II}) into (\ref{Casimir energy}), we derive the Casimir energy for JC I
\begin{eqnarray} \label{Casimir energy I}
  W_{\text{I}}=-\frac{\pi ^2+6 \left(\text{Li}_2\left(\frac{2 i \sqrt{2}}{3}-\frac{1}{3}\right)+\text{Li}_2\left(-\frac{2 i \sqrt{2}}{3}  -\frac{1}{3}\right)\right)}{4 \pi L}\approx -\frac{0.36}{L},
\end{eqnarray}
and for JC II
\begin{eqnarray} \label{Casimir energy II}
  W_{\text{II}}=-\frac{3 \left(\text{Li}_2\left(\frac{1}{3}+\frac{2 i \sqrt{2}}{3}\right)+\text{Li}_2\left(\frac{1}{3}-\frac{2 i \sqrt{2}}{3}\right)\right)}{2 \pi L}\approx -\frac{0.09}{L}.
\end{eqnarray}
The Casimir force is given by
\begin{eqnarray} \label{Casimir force}
  F=-\frac{1}{p}\frac{\partial W}{\partial L}=\frac{W}{p L},
\end{eqnarray}
with $p=6$ as the number of edges. 

We are interested in network binding energy (\ref{network binding energy}), which is defined as the difference in Casimir energy between the network and its isolated constituent units. This energy represents the minimum required to assemble a network from its parts and has potential applications in polymer networks. As discussed in the Introduction, we consider the isolated constituent units (strips) with identical boundary conditions. This results in a non-negative network binding energy due to the lower bound (\ref{sect 3: Casimir bound1}) on the network Casimir energy. Recall that the Casimir energy of a single strip is given by $W_{\text{strip}}=-\pi /(24 L)\approx -0.13/L$ for a free scalar. For the tetrahedron network, the binding energy can be expressed as follows. For JC I, we have 
\begin{eqnarray} \label{Casimir energy I}
W_{\text{bind\ I}}=W_{\text{I}}-6 W_{\text{strip}} \approx \frac{0.43}{L}.
\end{eqnarray}
And for JC II, we have
\begin{eqnarray} \label{Casimir energy I}
W_{\text{bind\ II}}=W_{\text{II}}-6 W_{\text{strip}} \approx \frac{0.70}{L}.
\end{eqnarray}

By using similar methods, we can calculate the Casimir energy and binding energy for various regular polyhedra. The results are summarized in Table \ref{table1} in the Introduction, and the corresponding spectra are provided in Appendix B. Both JC I and JC II produce an attractive Casimir force; however, JC II generally results in a smaller Casimir force compared to JC I. Notably, for the Hexahedron network, both JCs yield the same Casimir force. It is important to note that the positive definiteness of the binding energy is based on the lower bound described in (\ref{sect 3: Casimir bound1}) of the network Casimir energy. Table \ref{table1} shows that all the binding energy densities, denoted as $\rho_{\text{bind}}$, are positive for regular polyhedra. It strongly supports the argument that the lower bound stated in (\ref{sect 3: Casimir bound1}) applies to general networks, not just the simplest network with only one node.

\section{Conclusions and Discussions}

% v2 This paper explores the general junction conditions (JCs) and the Casimir effect for conformal field theories in networks (NCFT). For the sake of simplicity, we focus on the case of a free massless scalar field in $(1+1)$ dimensions.
This paper explores the general junction conditions (JCs) and the Casimir effect for $(1+1)$-dimensional free scalars in networks.
We begin by examining two typical JCs using the action principle and discuss how they can be implemented in real physical systems consisting of strings and thin rigid rods. Next, we address the most general JCs that satisfy energy conservation at the node. These general JCs are expressed through the $O(p)$ symmetry group, where $p$ is the number of edges linked at the node. Utilizing these general JCs, we derive both lower and upper bounds for the Casimir energy of the scalar field in the simplest network featuring only one node. 
Interestingly, these bounds correspond to the JCs with ideal reflection at nodes, where a wave cannot travel from one edge to another. Under these conditions, NCFTs behave as isolated BCFTs at each edge. The Casimir effect is inherently a boundary phenomenon; thus, the greater the reflection at the node, the stronger the Casimir effect.
 Thus, it is expected that the bounds of Casimir energy are given by JCs with ideal reflection for general NCFTs, not limited to the scalar in the simplest network. Recent works \cite{Miao:2024gcq, Miao:2025utb} suggest that there exists a lower bound on the Casimir effect for general BCFTs. This conjecture has been validated for $(1+1)$-dimensional BCFTs and has been tested successfully in higher dimensions. Motivated by these findings and the arguments presented above, we establish a lower bound for the network Casimir energy in general $(1+1)$-dimensional NCFTs. Our scalar example also indicates an upper bound on the network Casimir effect, which we plan to examine further in future research.

Additionally, we investigate the Casimir effect within networks of regular polyhedra under the two typical JCs. Our findings reveal that JC II typically yields a smaller Casimir force than JC I. Moreover, the network binding energy, defined as the difference in Casimir energy between NCFTs and isolated BCFTs, is found to be positive for all networks of regular polyhedra. It implies that energy is required to assemble a regular polyhedron network from its individual components (strips with identical boundary conditions). In the future, it would be interesting to explore the entanglement entropy in NCFTs. Holographic theory suggests that the difference in entanglement entropy between NCFTs and BCFTs is non-negative \cite{Guo:2025sbm}, i.e, $S_{\text{NCFT}}-S_{\text{BCFT}}\ge 0$. It would be intriguing to determine whether this holds for free theories as well. Additionally, it is worth considering extending the results of this paper to broader theories, particularly gravitational theories, which have significant implications for wormhole physics \cite{Shen:2024itl}, holographic networks \cite{Guo:2025sbm}, and traversable parallel universes \cite{Guo:2026prm}.

\section*{Acknowledgements}
% Zhao is supported by Guangdong Basic and Applied Basic Research Foundation (No.2021A1515010276) and National Natural Science Foundation of China (NSFC) grant (No.11904422). 
Miao acknowledges the supports from NSFC (No.12275366).

\section*{Appendix A}

In this appendix, we extend our discussion of scalar junction conditions (JCs) to higher dimensions and to Maxwell fields.

Taking variations of the scalar action yields boundary terms at the node \(N\):
\begin{eqnarray} \label{appB: action1}
  \delta I|_N= \int_{N} d^{d-1}y  \Big( \sum_{i=1}^p \partial_{n} \phi_i \delta \phi_i \Big)=0,
\end{eqnarray}
where $n$ is the normal direction to the node $N$, $\phi_i$ denotes the scalar field in edge $E_i$, $p$ is the number of edges and $d$ is the spacetime dimension.  We have two natural choices of JCs that ensure a well-defined variational principle, \(\delta I|_N=0\):
\begin{eqnarray} \label{appB: JC I}
 \text{JC I}:\  \lambda_i \phi_i|_N=\lambda_j \phi_j|_N,\ \ \sum_{i=1}^p \frac{1}{\lambda_i} \partial_n \phi_i |_N=0,
\end{eqnarray}
and
\begin{eqnarray} \label{appB: JC II}
 \text{JC II}:\   \lambda_i \partial_n \phi_i|_N=\lambda_j \partial_n \phi_j|_N,\ \ \sum_{i=1}^p \frac{1}{\lambda_i} \phi_i|_N=0,
\end{eqnarray} 
where $ \lambda_i $ are arbitrary constants. 

Recall that the general JCs for (1+1)-dimensional scalars are characterized by an $O(p)$ rotation group, which can be derived from the energy conservation at nodes
\begin{align}\label{appB: conservation law 3}
    \sum_{i}^{p}\overset{(i)}{T}_{nt}|_{N}=\sum_{i}^{p}\partial_{n}\phi_i \partial_{t}\phi_i|_{N}=0.
\end{align}
As discussed in Section 2.2, the junction conditions (\ref{appB: JC I}) and (\ref{appB: JC II}) correspond to the special rotation angle \(\theta = \pi\) in (1+1) dimensions. In higher dimensions, the scalar JCs are no longer characterized by the \(O(p)\) group due to additional constraints at the node. In addition to energy conservation (\ref{appB: conservation law 3}), we also require the conservation of tangential momentum flux at the node \cite{Guo:2025sbm}:
\begin{align}\label{appB: conservation law 4}
    \sum_{i}^{p}\overset{(i)}{T}_{na}|_{N}=\sum_{i}^{p}\partial_{n}\phi_i \partial_{a}\phi_i|_{N}=0,
\end{align}
where \(a\) labels the tangential direction to the node \(N\). These additional constraints reduce the junction conditions from the \(O(p)\) group to those given by (\ref{appB: JC I}) and (\ref{appB: JC II}). Since higher-dimensional CFTs possess fewer symmetries than 2D CFTs, it is natural that the allowed JCs for higher-dimensional CFTs are less than that of 2D CFTs.

Let us go on to discuss the JCs for Maxwell theory in general dimensions. By taking variations and focusing on the boundary terms at the nodes, we obtain 
\begin{eqnarray} \label{appB: dI Maxwell}
  \delta I|_N= \int_{N} d^{d-1}y  \Big( \sum_{i=1}^p \overset{(i)}{F}_{n}{}^{a}\delta  \overset{(i)}{A}_a \Big)=0,
\end{eqnarray}
where \(\overset{(i)}{A}_a\) represents the Maxwell field on edge \(E_i\), and \(\overset{(i)}{F}_{na}\) refers to the normal-tangential components of the field strength on edge \(E_i\). Similar to the scalar case, two natural JCs ensure that \(\delta I|_N=0\):
\begin{eqnarray} \label{appB: JC I A}
 \text{JC I}:\  \lambda_i \overset{(i)}{F}_{ab}|_N=\lambda_j \overset{(j)}{F}_{ab}|_N,\ \ \sum_{i=1}^p \frac{1}{\lambda_i} \overset{(i)}{F}_{na}|_N=0,
\end{eqnarray}
and
\begin{eqnarray} \label{appB: JC II A}
 \text{JC II}:\   \lambda_i  \overset{(i)}{F}_{na}|_N=\lambda_j  \overset{(j)}{F}_{na}|_N,\ \ \sum_{i=1}^p \frac{1}{\lambda_i} \overset{(i)}{F}_{ab}|_N=0,
\end{eqnarray} 
where \(\lambda_i\) are arbitrary constants, with \(n\) and \(a\) denoting the normal and tangential directions, respectively.

One can easily verify that the above JCs satisfy the conservation of energy and the tangential momentum flux at the node:
\begin{eqnarray} \label{appB: conservation law Maxwell}
 \sum_{i}^{p}\overset{(i)}{T}_{nb}|_{N}=\sum_{i}^{p} \overset{(i)}{F}_{n}{}^{a}  \overset{(i)}{F}_{ab}|_{N}=0. 
\end{eqnarray} 
For JC I, we have 
\begin{eqnarray} \label{appB: conservation law Maxwell I}
 \sum_{i}^{p}\overset{(i)}{T}_{nb}|_{N}=\sum_{i}^{p} (\frac{1}{\lambda_i} \overset{(i)}{F}_{n}{}^{a}) (\lambda_i  \overset{(i)}{F}_{ab}) |_{N}=(\lambda_i  \overset{(i)}{F}_{ab} |_{N})
\sum_{i}^{p} \frac{1}{\lambda_i} \overset{(i)}{F}_{n}{}^{a} |_{N}=0,
 \end{eqnarray} 
where we have used the fact that \(\lambda_i \overset{(i)}{F}_{ab} |_{N}\) is independent of the index \(i\) for JC I. Similarly, for JC II, we have
\begin{eqnarray} \label{appB: conservation law Maxwell II}
 \sum_{i}^{p}\overset{(i)}{T}_{nb}|_{N}=\sum_{i}^{p} (\frac{1}{\lambda_i} \overset{(i)}{F}_{n}{}^{a}) (\lambda_i  \overset{(i)}{F}_{ab}) |_{N}= (\frac{1}{\lambda_i} \overset{(i)}{F}_{n}{}^{a})
\sum_{i}^{p} \lambda_i  \overset{(i)}{F}_{ab} |_{N}=0,
 \end{eqnarray}  
where we have noted that \(\lambda_i \overset{(i)}{F}_{n}{}^{a} |_{N}\) is independent of the index \(i\) for JC II.

\section*{Appendix B}

For the simplest network with three edges and node, the spectrums for the general JC (\ref{general JC solution2}) are given by 
\begin{align}\label{app: JC 1}
\Delta_{\text{DBC}}^+&=\Delta_{\text{NBC}}^-=2 \sin ^2(\alpha ) e^{i \left(L_2+L_3\right) \omega } \Big(\cos (\alpha ) \sin (\beta ) (\cos (\theta )-1)-\cos (\beta ) \sin (\theta )\Big)\nonumber\\
& \times \Big(\cos (\beta ) \sin (\theta ) \cos \left(L_1 \omega \right)+2 i \cos (\alpha ) \sin (\beta ) \sin ^2\left(\frac{\theta }{2}\right) \sin \left(L_1 \omega \right) \Big)\nonumber\\
&+2 i \sin ^2(\alpha ) e^{i \left(L_1+L_3\right) \omega } \Big(\cos (\alpha ) \cos (\beta ) (\cos (\theta )-1)+\sin (\beta ) \sin (\theta )\Big)\nonumber\\
&\times \Big(2 \cos (\alpha ) \cos (\beta ) \sin ^2\left(\frac{\theta }{2}\right) \sin \left(L_2 \omega \right)+i \sin (\beta ) \sin (\theta ) \cos \left(L_2 \omega \right)\Big)\nonumber\\
&+\Big(-\cos \left(L_3 \omega \right) \left(\sin ^2(\alpha ) \cos (\theta )+\cos ^2(\alpha )+1\right)+2 i \sin ^2(\alpha ) \sin ^2\left(\frac{\theta }{2}\right) \sin \left(L_3 \omega \right)\Big)\nonumber\\
& \times {\color{red}\Big[} e^{i \left(L_1+L_2\right) \omega } \left(\cos ^2(\alpha ) \sin ^2(\theta )-\sin ^4(\alpha ) \sin ^2(2 \beta ) \sin ^4\left(\frac{\theta }{2}\right)\right) \nonumber\\
& + \Big(\cos \left(L_1 \omega \right) \left(\sin ^2(\alpha ) \cos ^2(\beta ) (\cos (\theta )-1)-\cos (\theta )-1\right)+i (\cos (\theta )-1) \sin \left(L_1 \omega \right) \left(\sin ^2(\alpha ) \cos ^2(\beta )-1\right) \Big) \nonumber\\
& \times \Big(\cos \left(L_2 \omega \right) \left(\sin ^2(\alpha ) \sin ^2(\beta ) (\cos (\theta )-1)-\cos (\theta )-1\right)+i (\cos (\theta )-1) \left(\sin ^2(\alpha ) \sin ^2(\beta )-1\right) \sin \left(L_2 \omega \right) \Big)  {\color{red}\Big]} 
\end{align}
and 
\begin{align}\label{app: JC 2}
\Delta_{\text{DBC}}^-=&\Delta_{\text{NBC}}^+=4 \sin ^2(\alpha ) \sin ^2\left(\frac{\theta }{2}\right) \cos \left(L_3 \omega \right) \left(\cos (2 \beta ) \sin \left(\left(L_1-L_2\right) \omega \right)+\sin \left(\left(L_1+L_2\right) \omega \right)\right) \nonumber\\
&+\sin \left(L_3 \omega \right) \left(\cos \left(\left(L_1-L_2\right) \omega \right)-2 \left(\cos (2 \alpha )-2 \cos ^2(\alpha ) \cos (\theta )\right) \cos \left(L_1 \omega \right) \cos \left(L_2 \omega \right)\right) \nonumber\\
&+\sin \left(L_3 \omega \right) \left(4 \cos (\theta ) \sin \left(L_1 \omega \right) \sin \left(L_2 \omega \right)-3 \cos \left(\left(L_1+L_2\right) \omega \right)\right).
\end{align}
Here DBC and NBC are the boundary conditions (\ref{sect 3:  DBC NBC}) at the outer boundaries, $``\pm"$ indicates the choice of \( S \) from (\ref{sect2: S1}).

Following the approach of sect. 4, we can derive the spectra for various regular polyhedra. Substituting these spectrums into formulas (\ref{Casimir energy},\ref{Casimir force}), we can obtain the Casimir force and the data listed in the Table. \ref{table1}. The spectra for various regular polyhedra are listed below, where ``H, O, D, I" denote ``Hexahedron, Octahedron,  Dodecahedron, Icosahedron" respectively. 
\begin{eqnarray} \label{Casimir force H}
 &&\Delta^{\text{H}}_{\text{I}}(\omega)=\Delta^{\text{H}}_{\text{II}}(\omega)=\sin^6\left(L \omega \right)\left(7+9\cos \left(2 L \omega\right)\right)^3=0,\\ \label{Casimir force O I}
&&\Delta^{\text{O}}_{\text{I}}(\omega)=\sin^2\left(\frac{L \omega}{2}  \right)\sin\left(L \omega \right)\sin^3\left(2 L \omega \right) \left(\sin\left(L \omega \right)+\sin\left(2 L \omega \right)\right)^2=0,\\
&& \Delta^{\text{O}}_{\text{II}}(\omega)=\cos^2\left(\frac{3 L \omega}{2}  \right)\sin^3\left(L \omega \right)\sin^3\left(2 L \omega \right)=0, \label{Casimir force O II} \\
&& \Delta^{\text{D}}_{\text{I}}(\omega)= \cos^{10}\left(\frac{L \omega}{2}  \right)\cos^4\left(L \omega \right)\left(3 \cos\left(L \omega \right)-1 \right)^5 \left(2+3\cos\left(L \omega \right) \right)^4 \nonumber\\  \label{Casimir force D I}
&&\ \ \ \ \ \ \ \ \ \ \ \ \ \   \times \left(9\cos\left(2 L \omega \right)-1\right)^3 \sin^{12}\left(\frac{L \omega}{2} \right)= 0,\\
&&\Delta^{\text{D}}_{\text{II}}(\omega)=\cos^{12}\left(\frac{L \omega}{2} \right)\left(2-3\cos\left(L \omega \right) \right)^4 \cos^4\left(L \omega \right)\left(1+3 \cos\left(L \omega \right) \right)^5\nonumber\\
&&\ \ \ \ \ \ \ \ \ \ \ \ \ \  \times \left(9\cos\left(2 L \omega \right)-1\right)^3 \sin^{10}\left(\frac{L \omega}{2}  \right)=0, \label{Casimir force D II}\\
&&\Delta^{\text{I}}_{\text{I}}(\omega)=\cos^{18}\left(\frac{L \omega}{2}  \right)\left(1+5\cos\left(L \omega \right) \right)^5\left(3+5\cos\left(2 L \omega \right) \right)^3\sin^{20}\left(\frac{L \omega}{2}\right)=0,  \label{Casimir force I I}\\
&&\Delta^{\text{I}}_{\text{II}}(\omega)=\cos^{20}\left(\frac{L \omega}{2}\right)\left(-1+5\cos\left(L \omega \right) \right)^5\left(3+5\cos\left(2 L \omega \right) \right)^3\sin^{18}\left(\frac{L \omega}{2}  \right)=0. \label{Casimir force I II}
\end{eqnarray}

\end{document}